\documentclass[a4paper,11pt]{article}
\pdfoutput=1 

\usepackage{jcappub} 
\usepackage{indentfirst}
\usepackage{appendix}
\usepackage{multirow}

\title{Quantifying the impacts of future gravitational-wave data on constraining interacting dark energy}

\author[a]{Hai-Li Li,}
\author[a]{Dong-Ze He,}
\author[a]{Jing-Fei Zhang,}
\author[a,b,c,1]{Xin Zhang\note{Corresponding author.}}

\affiliation[a]{Department of Physics, College of Sciences, Northeastern
University, Shenyang 110819, China}
\affiliation[b]{Ministry of Education's Key Laboratory of Data Analytics and Optimization
for Smart Industry, Northeastern University, Shenyang 110819, China}
\affiliation[c]{Center for High Energy Physics, Peking University, Beijing 100080, China}

\emailAdd{1329750467@qq.com, hedongze1992@163.com, jfzhang@mail.neu.edu.cn, zhangxin@mail.neu.edu.cn}

\abstract{In this work, we investigate the impacts of the future gravitational-wave (GW) standard siren observation by the Einstein Telescope (ET) on constraining the interacting dark energy (IDE) models. We simulate 1000 GW events in the redshift range of $0\lesssim z \lesssim 5$ based on the 10-year observation of the ET. We combine the simulated GW data with the current mainstream cosmological electromagnetic observations including the cosmic microwave background anisotropies, the baryon acoustic oscillations, and the type Ia supernovae to constrain the IDE models. We consider typical IDE models in the context of a perturbed universe. To avoid the large-scale instability problem for IDE models, we apply the extended parameterized post-Friedmann approach to calculate the cosmological perturbations. We find that the addition of the GW standard siren data could significantly improve the constraint accuracies for most of the cosmological parameters (e.g., $H_{0}$, $w$, and $\Omega_{\rm m}$). For the coupling parameter $\beta$, the constraint errors could also be slightly improved when adding the GW data in the cosmological fit.}

\begin{document}
\maketitle
\flushbottom

\section{Introduction}
\label{sec1}

The accelerated expansion of the universe, discovered by the observations of type Ia supernovae \cite{Riess:1998cb,Perlmutter:1998np} and further confirmed by the observations of cosmic microwave background \cite{Spergel:2003cb,Bennett:2003bz} and large scale structure \cite{Tegmark:2003ud,Abazajian:2004aja}, has become a fact.
In order to explain the cosmic acceleration, the concept of ``dark energy'',  which is an exotic form of energy with negative pressure, has been proposed \cite{Sahni:2006pa,Bamba:2012cp,Weinberg:1988cp,Peebles:2002gy,Copeland:2006wr,Frieman:2008sn,Sahni:2008zz,Li:2011sd,Kamionkowski:2007wv}. At present, dark energy (DE) occupies about 68\% of the total energy density of the cosmos, dominating the evolution of the current universe.

The cosmological constant $\Lambda$, proposed by Einstein in 1917, has always been regarded as the simplest candidate of DE until now. The combination of cosmological constant $\Lambda$ (or vacuum energy) and cold dark matter (CDM) constitute a concordant cosmological model, called the $\Lambda$CDM model. The equation of state (EoS) parameter of vacuum energy is $w_{\Lambda} \equiv p_{\Lambda}/\rho_{\Lambda}=-1$. Although the $\Lambda$CDM model is in excellent agreement with current cosmological observations with the least parameters \cite{Ade:2015xua}, the cosmological constant $\Lambda$ has always been plagued by some severe theoretical puzzles, such as the ``fine-tuning'' and ``cosmic coincidence'' problems \cite{Sahni:1999gb,Bean:2005ru}. Thus, it is hard to say that the cosmological constant model with only six free parameters is the eventual scenario of our universe, which implies that the $\Lambda$CDM model is necessary to be further extended and some new parameters concerning new physics are thus introduced into the extensions.

To extend the $\Lambda$CDM cosmology in the aspect of DE, there are mainly two possible theoretical orientations, i.e., dynamical dark energy and modified gravity (MG) theories. If Einstein's general relativity (GR) is valid on all the scales of universe, an alternative proposal of $\Lambda$ is the dynamical dark energy, which suggests that the energy form with negative pressure can be provided by a spatially homogeneous scalar field evolving slowly down a proper potential, dubbed quintessence. On the other hand, if GR breaks down on the cosmological scales, some models of MG can mimic the ``effective dark energy" at the cosmological background level to explain the cosmic accelerated expansion. In general, a dynamical dark energy model, compared to the cosmological constant, can yield a different expansion history of the universe but a similar growth history of structure. On the contrary, the MG models can yield a similar expansion history but a quite different structure growth history. Discriminating the scenarios of dynamical dark energy and MG has become one of the most critical issues in modern cosmology.

However, it must be emphasized that there is another important theoretical possibility that dark energy and dark matter can directly interact with each other, through mediating some unknown scalar field degrees of freedom, also called ``the fifth force". Inspired by this possibility, a large number of models featuring the interaction between dark matter and dark energy have been constructed and researched \cite{Amendola:1999er,Amendola:1999qq,TocchiniValentini:2001ty,Amendola:2001rc,Comelli:2003cv,Chimento:2003iea,Cai:2004dk,Zhang:2004gc,Ferrer:2004nv,Zimdahl:2005bk,Zhang:2005rj,Wang:2006qw,Sadjadi:2006qp,Barrow:2006hia,Sasaki:2006kq,Abdalla:2007rd,Bean:2007ny,Guo:2007zk,Bertolami:2007zm,Boehmer:2008av,He:2008tn,CalderaCabral:2008bx,Bean:2008ac,Szydlowski:2008by,Chen:2008ft,Valiviita:2008iv,Couderc:2009tq,Chimento:2009hj,CalderaCabral:2009ja,Majerotto:2009np,Valiviita:2009nu,He:2009mz,He:2009pd,Koyama:2009gd,Li:2009zs,Xia:2009zzb,Cai:2009ht,He:2010ta,Cui:2010dr,Li:2010eu,Gavela:2010tm,Martinelli:2010rt,He:2010im,Chen:2011rz,Fu:2011ab,Clemson:2011an,Li:2011ga,Xu:2011tsa,Zhang:2012uu,Xu:2013jma,Zhang:2013zyn,Wang:2013qy,Salvatelli:2013wra,Yang:2014gza,yang:2014vza,Wang:2014oga,Faraoni:2014vra,Yin:2015pqa,Fan:2015rha,Cai:2015zoa,Duniya:2015nva,Feng:2016djj,Murgia:2016ccp,Sola:2016jky,Sola:2016ecz,Sola:2016zeg,Pourtsidou:2016ico,Costa:2016tpb,Xia:2016vnp,vandeBruck:2016hpz,Kumar:2016zpg,Kumar:2017dnp,Santos:2017bqm,Sola:2017jbl,Guo:2017hea,Zhang:2017ize,Feng:2018yew,Yang:2018euj,Zhao:2018fjj,Li:2019loh}.
Although the interaction between dark matter and dark energy is mild, we still cannot exclude it within $1\sigma$ confidence region \cite{Costa:2013sva,Salvatelli:2014zta,Nunes:2016dlj, Abdalla:2014cla,Yang:2017ccc,Yang:2017zjs,Li:2018ydj,Feng:2019mym}. What is important is that the models of interacting dark energy can successfully solve  (or alleviate) the cosmic coincidence problem through the attractor solution. Recently, the models of interacting dark energy have been considered to help alleviate the Hubble constant tension between the early and late universe measurements \cite{Guo:2018ans}, and it was also shown that they are helpful in partially explaining the excess of 21 cm absorption signal related to the epoch of cosmic dawn (at $z \sim 17$) detected in the EDGES experiment  \cite{Costa:2018aoy,Xiao:2018jyl,Li:2019loh}. Thus, the research on interacting dark energy models is expected to be significant and valuable.


Currently, the major cosmological probes include the cosmic microwave background (CMB) anisotropies, baryon acoustic oscillations (BAO), type Ia supernovae (SN), direct determination of the Hubble constant ($H_{0}$), weak gravitational lensing (WL), redshift space distortions (RSD), etc. Some important cosmological parameters have been precisely measured by the combination of these electromagnetic (EM) probes. But for the parameters beyond the standard model, such as the EoS of dark energy, the sterile neutrino mass, the tensor-to-scalar ratio and so forth, we still cannot measure them accurately up to now. In fact, there are strong degeneracies between these parameters, and some conflicts also exist among various observations. The reason for the situation is that the current observations are still not accurate enough, so that we cannot precisely measure the cosmological parameters beyond the standard model. In order to better constrain these parameters, we also need some new cosmological probes other than the traditional EM cosmological probes.

As proposed by Schutz in 1986 \cite{Schutz:1986gp} and subsequently discussed by Holz and Hughes \cite{Holz:2005df}, the observations of gravitational waves (GW) can be used as the \textit{standard sirens} in cosmology.
The detection of GW event GW170817 \cite{TheLIGOScientific:2017qsa} from the merger of binary neutron star and its EM counterpart GRB170817A \cite{Monitor:2017mdv} have pronounced the arrival of the multi-messenger astronomy era. With the help of the multi-messenger observation, we can measure the absolute luminosity distance $d_{\rm L}$ of the source from the gravitational wave signal as well as the redshift $z$ from the observation for its electromagnetic counterpart. Then, we can establish a true distance-redshift relation which can be used to infer the expansion history of universe and constrain the cosmological parameters such as the Hubble constant \cite{Abbott:2017xzu}. For measuring the Hubble constant, the main advantage of the standard siren method is that it avoids using the cosmic distance ladder. The GW standard sirens would become a promising new cosmological probe in the future, and would play a significant role in the cosmological parameter measurements \cite{Zhang:2019ylr,Xu:2020uws}.

Actually, in the 2030s, a proposed third-generation ground-based GW observatory, the Einstein Telescope (ET), will be brought into operation \cite{ET}. This impressive facility will have 10 km-long arms and three detectors. Compared with the advanced LIGO, it has a much wider detection frequency range and a much better detection sensitivity. Thus, there will be much more binary neutron star merger events in much deeper redshifts detected by ET. As a conservative estimation, at least 1000 useful standard siren events will be observed with ET's ten-year operation \cite{Zhang:2018byx}.
In the literature \cite{Zhao:2018gwk,Du:2018tia,Zhang:2018byx,Zhang:2019ple,Yang:2019bpr,Yang:2019vni,Bachega:2019fki,Zhang:2019loq,Zhang:2019ylr,Chang:2019xcb,He:2019dhl,Liu:2017xef,Berti:2018cxi,Liu:2018sia,Will:1994fb,Wang:2019tto,Zhao:2019gyk}, some authors have utilized the simulated future GW standard siren observations to estimate the cosmological parameters in various cosmological models of dark energy. For example, in Ref.~\cite{Zhang:2018byx} the authors have investigated the capabilities of future GW standard siren observation for improving the parameter estimation in cosmology and for breaking the parameter degeneracies formed in traditional EM observations. Taking ET as an example, they simulated 1000 data based on the ten-year observation and found that the simulated GW data could effectively break the parameter degeneracy in the $\Lambda$CDM and $w$CDM models, significantly improving the parameter constraints in the cosmological fit. In Ref. \cite{Zhang:2019loq}, the authors have also investigated the Chevalliear-Polarski-Linder (CPL), $\alpha$ dark energy ($\alpha$DE), generalized Chaplygin gas (GCG) and new generalized Chaplygin gas (NGCG) models with the simulated GW standard siren data, and it was shown that the GW data could also improve the constraints on the cosmological parameters for all these DE models. Likewise, the similar conclusion can also be drawn for the holographic dark energy models, as shown in Ref.~\cite{Zhang:2019ple}. In addition to the DE models, the GW standard siren observations can also exert significant influences on the studies of the MG models. The impacts of the future GW observations on the MG models have been recently intensively discussed in the literature (e.g., Refs.~\cite{Will:1994fb,Liu:2017xef,Berti:2018cxi,Zhao:2018gwk,Liu:2018sia}).

As for the interacting dark energy (IDE) models, Ref.~\cite{Yang:2019vni} has recently investigated how the future GW data could help improve the limits on the parameters in two specific I$\Lambda$CDM models, finding that the addition of GW data to the CMB data can reduce the current uncertainty by a factor of 5. However, it is well-known that for the IDE scenario there is a problem of early-time perturbation instability, and thus in Ref.~\cite{Yang:2019vni} the authors have to set the coupling constant $\xi$ to be positive and introduce a factor (1+$w_{x}$) into the interaction term $Q$ to ensure the stability for these models. In the present work, to treat the large-scale instability problem for the cosmological perturbations in the IDE scenario, we adopt the extended parameterized post-Friedmann (PPF) method \cite{Li:2014eha,Li:2014cee,Li:2015vla,Zhang:2017ize,Feng:2018yew,Feng:2019mym,Feng:2019jqa} to calculate the cosmological perturbations. Using the extended PPF method, without assuming any specific ranges of the EoS parameter $w$ and the coupling constant $\beta$, the cosmological perturbations can be calculated safely in the whole parameter space of an interacting dark energy model.
We will further explore the impacts of the future GW data on improving the parameter constraints by breaking the parameter degeneracies for the IDE models.
We consider two cases of interaction term, i.e., $Q=\beta H \rho_{\rm c}$ and $Q=\beta H_{0} \rho_{\rm c}$, where $\rho_{\rm c}$ is the energy density of cold dark matter. 
This work will make the analysis of impacts of GW standard sirens on improving cosmological parameter estimation more complete.


This paper is organized as follows. In Sec.~\ref{sec2}, we give a brief description for the extended PPF approach to the interacting dark energy models. In Sec.~\ref{sec3}, we introduce the current cosmological data and briefly describe the method to simulate the GW data. In Sec.~\ref{sec4}, we report the constraint results and make some relevant discussions. Conclusion is given in Sec.~\ref{sec5}.

\section{A brief description of the extended PPF approach for interacting dark energy models}\label{sec2}

If there is a direct, non-gravitational interaction between dark energy and dark matter, we will have the following energy continuity equations
\begin{align}
&\rho'_{\rm de} = -3\mathcal{H}(1+w)\rho_{\rm de}+ aQ_{\rm de},\label{eq1}\\
&\rho'_{\rm c} = -3\mathcal{H}\rho_{\rm c}+aQ_{\rm c},~~~~~~Q_{\rm de}=-Q_{\rm c}=Q,\label{eq2}
\end{align}
where $\rho_{\rm de}$ and $\rho_{\rm c}$ are the energy densities of dark energy and dark matter, respectively, a prime is the derivative with respect to the conformal time $\eta$, $\mathcal{H}=a'/a$ is the conformal Hubble expansion rate, $a$ denotes the scale factor, $w$ is the EoS parameter, and $Q$ is the phenomenological interaction term. Generally, the form of $Q$ is assumed to be proportional to the density of dark sectors, and it can include the Hubble parameter $H$ or the Hubble constant $H_{0}$. In this paper, we consider two forms of the interaction term $Q$, e.g., $Q=\beta H \rho_{\rm c}$ and $Q=\beta H_{0} \rho_{\rm c}$, with $\beta$ being a dimensionless coupling parameter used to describe the interaction strength between dark energy and dark matter. From Eqs. (\ref{eq1}) and (\ref{eq2}), we can clearly see that if $\beta>0$, dark matter would decay into dark energy, and vice versa for $\beta<0$. Here, $\beta=0$ denotes no interaction between the two sectors.

The covariant conservation law of the dark sectors can be expressed as
\begin{equation}
\label{eqn:energyexchange} \nabla_\nu T^{\mu\nu}_I = Q^\mu_I, \quad\quad
 \sum_I Q^\mu_I = 0,
\end{equation}
where $T^{\mu\nu}_I$ is the energy-momentum tensor, and $Q^\mu_I$ is the energy-momentum transfer vector. In this paper, we choose $Q^\mu_{\rm de}=-Q^\mu_{\rm c}=Qu^\mu_{\rm c}$, where $u^\mu_{\rm c}$ is the four-velocity of dark matter. The energy-momentum transfer vector $Q^\mu_I$ can be split into two parts as
\begin{equation}
Q^I_\mu  = a\big( -Q_I(1+AY) - \delta Q_IY,\,[ f_I+ Q_I (v-B)]Y_i\big),\label{eq:Qenergy}
\end{equation}
where $\delta Q_I$ is the energy transfer perturbation and $f_I$ is the momentum transfer potential of the $I$ fluid. $A$ and $B$ are the scalar metric perturbations. $Y$ is the eigenfunctions of the Laplace operator ($\nabla^{2}Y=-k^{2}Y$) and $Y_i$ is the covariant derivative ($Y_{i}=(-k)\nabla_{i}Y$).

In the interacting dark energy models, we can give the following conservation equations for the $I$ fluid according to Eqs.~(\ref{eqn:energyexchange}) and (\ref{eq:Qenergy}),
\begin{equation}
 {\delta\rho_I'}
	+  3\mathcal{H}({\delta \rho_I}+ {\delta p_I})+(\rho_I+p_I)(k{v}_I + 3 H_L')=a(\delta Q_I+AQ_I),\label{eqn:conservation1}
\end{equation}	
\begin{align}	
[(\rho_{I}+p_{I})(v_{I}-B)]'+4\mathcal{H}(\rho_I + p_I)({{v_I}-{B}})-k{ \delta p_I }
+\frac{2}{3}kc_{K} p_I {\Pi_I}- k(\rho_I+ p_I) {A}\nonumber \\
=a[Q_I(v-B)+f_I].\label{eqn:conservation2}
\end{align}

In the equations above, $\delta\rho_I$ is the energy density perturbation, $\delta p_I$ is the isotropic pressure perturbation, $v_I$ is the velocity perturbation, and $c_K = 1-3K/k^2$ with $K$ being the spatial curvature, and $\Pi_I$ is the anisotropic stress perturbation.

When considering the interaction between dark matter and dark energy, dark energy is treated as a nonadiabatic fluid and the calculation of $\delta p_{\rm de}$ is in terms of the adiabatic sound speed and the rest-frame sound speed. Under such circumstances, the interacting dark energy models will be plagued with the problem of large-scale instability. Hence, we should treat the dark energy perturbations with the extended PPF framework~\cite{Li:2014eha}. For clarity, in the following discussion, we will use some new symbols, i.e., $\zeta\equiv H_L$, $\xi\equiv A$, $\rho\Delta\equiv\delta\rho$, $\Delta p\equiv\delta p$, $V\equiv v$, and $\Delta Q_I\equiv\delta Q_I$, to denote the corresponding quantities of the comoving gauge, except the two gauge-independent quantities $\Pi$ and $f_I$.

On the large scales, the direct relationship between $V_{\rm de} - V_T$ and $V_T$ can be established, and it can be parametrized by a function $f_\zeta(a)$ as \cite{Hu:2008zd,Fang:2008sn}
\begin{equation}
\lim_{k_H \ll 1}\frac{4\pi G a^2}{\mathcal{H}^2}(\rho_{\rm de} + p_{\rm de})\frac{V_{\rm de} - V_T}{k_H}=-\frac{1}{3}ck_{K} f_\zeta(a) k_H V_T,\label{eq:DEcondition}
\end{equation}
where $k_H=k/\mathcal{H}$. The equation of motion for the curvature perturbation $\zeta$ on the large scales can be obtained by combining this equation with Einstein equations,
\begin{align}
\lim_{k_H \ll 1} \zeta'  = \mathcal{H}\xi - \frac{K}{k} V_T +\frac{1}{3} ck_{K}  f_\zeta(a) k V_T.
\label{eqn:zetaprimesh}
\end{align}
On the small scales, one can describe the evolution of curvature perturbation by using the Poisson equation, $\Phi=4\pi G a^2\Delta_T \rho_T/( k^2 c_{K})$, with $\Phi=\zeta+V_T/k_H$. These two limits can be linked by the introduction of a dynamical function $\Gamma$,
\begin{equation}
\Phi+\Gamma = {4\pi Ga^2
\over  k^2 c_{K}} \Delta_T \rho_T,
\label{eqn:modpoiss}
\end{equation}
which is satisfied for all the scales.

Compared with the small-scale Poisson equation, Eq.~(\ref{eqn:modpoiss}) gives $\Gamma\rightarrow0$ at $k_H\gg1$. Combining the derivative of Eq.~(\ref{eqn:modpoiss}) with the conservation equations and the Einstein equations, the equation of motion for $\Gamma$ on the large scales can be expressed as follows,
\begin{equation}\label{eq:gammadot}
\lim_{k_H \ll 1} \Gamma'  = S -\mathcal{H}\Gamma,
\end{equation}
with
\begin{align}
S&={4\pi Ga^2
\over k^2 } \Big\{[(\rho_{\rm de}+p_{\rm de})-f_{\zeta}(\rho_T+p_T)]kV_T \nonumber\\
&\quad+{3a\over k_Hc_K}[Q_{\rm c}(V-V_T)+f_{\rm c}]+\frac{a}{c_K}(\Delta Q_{\rm c}+\xi Q_{\rm c})\Big\},\nonumber
\end{align}
where $\xi$ can be obtained from Eq.~(\ref{eqn:conservation2}),
\begin{equation}
\xi =  -{\Delta p_T - {2\over 3}\prod_{T}+{a\over k}[Q_{\rm c}(V-V_T)+f_{\rm c}] \over \rho_T + p_T}.
\label{eqn:xieom}
\end{equation}

By the transition scale parameter $c_\Gamma$, we can take the equation of motion for $\Gamma$ on all scales to be \cite{Hu:2008zd,Fang:2008sn}
\begin{equation}
(1 + c_\Gamma^2 k_H^2) [\Gamma' +\mathcal{H} \Gamma + c_\Gamma^2 k_H^2 \mathcal{H}\Gamma] = S.
\label{eqn:gammaeom}
\end{equation}

From Eq. (\ref{eqn:gammaeom}) we can see that in the equation of motion for $\Gamma$, all of the perturbation quantities contain only matters and do not include dark energy. So, we can also solve Eq. (\ref{eqn:gammaeom}) without using any information related to the dark energy perturbations. As long as we know the evolution of $\Gamma$, we can get the energy density and velocity perturbations immediately,
\begin{align}
\rho_{\rm de}\Delta_{\rm de} =- 3(\rho_{\rm de}+p_{\rm de}) {V_{\rm de}-V_{T}\over k_{H} }-{k^{2} c_{K} \over 4\pi G a^{2}} \Gamma,\\
 V_{\rm de}-V_{T} ={-k \over 4\pi Ga^2 (\rho_{\rm de} + p_{\rm de}) F} \nonumber\\
\quad\quad\quad\times\left[ S - \Gamma' - \mathcal{H}\Gamma + f_{\zeta}{4\pi Ga^2 (\rho_{T}+p_{T}) \over k}V_{T}
\right],
\end{align}
with $F = 1 +  12 \pi G a^2 (\rho_T + p_T)/( k^2 c_{K})$.

\section{Data and method}\label{sec3}

In this section we will first describe the current observational data used in this paper, and then introduce the simulated future GW data.

The current observational data sets we used in this work include CMB, BAO and SN. For the CMB data, we use the Planck temperature and polarization power spectra of the full range of multipoles \cite{Aghanim:2015xee}, denoted here as ``Planck TT, TE, EE+lowTEB''. For the BAO data, we use the measurements from 6dFGS ($z_{\rm eff}=0.106$) \cite{Beutler:2011hx}, SDSS-MGS ($z_{\rm eff}=0.15$) \cite{Ross:2014qpa}, and BOSS DR12 ($z_{\rm eff}=0.38$, 0.51, and 0.61) \cite{Alam:2016hwk}. For the SN data, we use the Pantheon sample, which is comprised of 1048 data points \cite{Scolnic:2017caz}.

Next, we shall introduce the method to generate the GW standard siren data specifically. Each data point  consists of a triple ($z_{i}$, $d_{L}(z_{i})$, $\sigma_{i}$). Here, $z_{i}$ is the redshift of the GW source, $d_{L}(z_{i})$ is the luminosity distance at $z_{i}$, and $\sigma_{i}$ is the error. The simulation method is the same as described in Refs.~\cite{Cai:2016sby,Zhao:2010sz,Wang:2018lun,Zhang:2018byx}. The GW sources considered in this work include the merger events from black hole-neutron star (BHNS) systems and binary neutron star (BNS) systems, both are expected to exhibit afterglows in the electromagnetic (EM) radiation after they emit a burst of GW. Thus, BNS and NSBH could be observed as not only a transient GW event, but also an EM counterpart. The consideration of NSBH is likely to exert a beneficial effect on the determination of cosmological parameters \cite{Zhao:2010sz}.
But for NSBH, the intrinsic coalescence rate is expected to be considerably lower than that for BNS as indicated in the Conceptual Design Study of the Einstein Telescope \cite{ET} (see Table 2 on Page 31). Thus, for the GW standard siren data simulation, we mainly consider the coalescence events of BNS and only consider a small number of NSBH coalescence events. According to the prediction of the Advanced LIGO-Virgo network \cite{Abadie:2010px}, the radio between NSBH and BNS is set to be 0.03 so as to make BNS the majority of GW sources for standard sirens, which is also in accordance with Refs.~\cite{Cai:2016sby,Zhang:2019loq,Zhang:2018byx,Wang:2018lun,Zhang:2019ple}.

The redshift distribution of the observable sources is given by \cite{Cai:2016sby,Zhao:2010sz}
\begin{equation}
P(z)\propto \frac{4\pi d_C^2(z)R(z)}{H(z)(1+z)},
\label{equa:pz}
\end{equation}
where $d_C(z)$ is the comoving distance at the redshift $z$, and $R(z)$ denotes the redshift evolution of the burst rate that takes the form as  \cite{Schneider:2000sg,Cutler:2009qv,Cai:2016sby}
\begin{equation}
R(z)=\begin{cases}
1+2z, & z\leq 1, \\
\frac{3}{4}(5-z), & 1<z<5, \\
0, & z\geq 5.
\end{cases}
\label{equa:rz}
\end{equation}

Furthermore, we can get the catalogue of the GW sources by choosing a fiducial model. Theoretically, the fiducial model could be any well motivated cosmological model.
In this paper, we take the best-fit interacting dark energy models (i.e., the I$\Lambda$CDM model and the I$w$CDM model) constrained by the current observations as the fiducial models to produce the simulated GW data. For the base 6-parameter $\Lambda$CDM model, the cosmological parameters are $\{\omega_b,~\omega_c,~100\theta_{\rm MC},~\tau,~n_s,~\ln (10^{10}A_s)\},$ where $\omega_b=\Omega_bh^2$ and $\omega_c=\Omega_ch^2$ are the present densities of baryons and cold dark matter, respectively, $\theta_{\rm MC}$ is the ratio between the sound horizon and the angular diameter distance at the decoupling epoch, $\tau$ is the reionization optical depth, $n_s$ is the scalar spectral index, and $A_s$ is the amplitude of primordial scalar perturbation power spectrum. As an extension of the $\Lambda$CDM model, the $w$CDM model has an additional parameter $w=\rm constant$ relative to the $\Lambda$CDM model. Similarly, the I$\Lambda$CDM model has an additional coupling parameter $\beta$ relative to the $\Lambda$CDM model, and the I$w$CDM model has an additional coupling parameter $\beta$ relative to the $w$CDM model.

The comoving distance $d_C(z)$ can be calculated by the function
\begin{equation}
{d_C}(z) = \frac{{1}}{{H_0}}\int_0^z {\frac{{dz'}}{{E(z')}}},
\label{equa:dl}
\end{equation}
where $E(z)=H(z)/H_0$ is given by a cosmological model. Therefore, according to the redshift distribution of the GW sources, we can generate a catalog of the GW sources by Eq.~(\ref{equa:dl}), which means that the relation between $z$ and $d_L$ can be given for each fiducial model.

Since the GW amplitude depends on the luminosity distance $d_L$, the information of $d_L$ and $\sigma_{d_L}$ can be obtained from the waveform. The strain $h(t)$ in the GW interferometers can be written as
\begin{equation}
h(t)=F_+(\theta, \phi, \psi)h_+(t)+F_\times(\theta, \phi, \psi)h_\times(t),
\end{equation}
where $F_{+}$ and $F_{\times}$ are the beam pattern functions, $\psi$ is the polarization angle, $\theta$ and $\phi$ describe the location of the GW source relative to the GW detector. The antenna pattern functions of the ET can be written as \cite{Zhao:2010sz}
 \begin{align}
F_+^{(1)}(\theta, \phi, \psi)=&~~\frac{{\sqrt 3 }}{2}[\frac{1}{2}(1 + {\cos ^2}(\theta ))\cos (2\phi )\cos (2\psi ) \nonumber\\
                              &~~- \cos (\theta )\sin (2\phi )\sin (2\psi )],\nonumber\\
F_\times^{(1)}(\theta, \phi, \psi)=&~~\frac{{\sqrt 3 }}{2}[\frac{1}{2}(1 + {\cos ^2}(\theta ))\cos (2\phi )\sin (2\psi ) \nonumber\\
                              &~~+ \cos (\theta )\sin (2\phi )\cos (2\psi )].
\label{equa:F}
\end{align}
Obviously, the antenna pattern functions of the other two interferometers can also be easily calculated due to fact that the three interferometers are placed in an equilateral triangle shape, with the angles with each other being $60^{\circ}$.

Next, we compute the Fourier transform $\mathcal{H}(f)$ of the time domain waveform $h(t)$,
\begin{equation}
\mathcal{H}(f)=\mathcal{A}f^{-7/6}\exp[i(2\pi ft_0-\pi/4+2{\Psi}(f/2)-\varphi_{(2.0)})].
\label{equa:hf}
\end{equation}
{Here, the functions $\Psi(f)$ and $\varphi_{(2.0)}$ are
\begin{equation}\label{psi}
\Psi(f)=-\psi_{0}+\frac{3}{256\eta}\sum_{i=0}^7\psi_i(2\pi
Mf)^{i/3},
\end{equation}
\begin{equation}\label{phi}
\varphi_{(2,0)}=
\tan^{-1}\left(-\frac{2\cos(\iota)F_{\times}}{(1+\cos^2(\iota))F_{+}}\right),
\end{equation}
with the parameters $\psi_i$ provided in Ref. \cite{Sathyaprakash:2009xs}}. The Fourier amplitude $\mathcal{A}$ can be expressed as
\begin{equation}
\mathcal{A}=\frac{1}{d_L}\sqrt{F_+^2(1+\cos^2(\iota))^2+4F_\times^2\cos^2(\iota)}
             \sqrt{5\pi/96}\pi^{-7/6}\mathcal{M}_c^{5/6},
\label{equa:A}
\end{equation}
where $\mathcal{M}_c=M \eta^{3/5}$ is the ``chirp mass" related to the total mass $M$ of the coalescing binary system, $\eta=m_1 m_2/M^2$ is the symmetric mass ratio, and $m_1$ and $m_2$ are the component masses. 
The masses quoted here refer to the redshifted masses in observation, and the relation between the observed mass and the intrinsic mass (in the source frame) is $M_{\rm obs}=(1+z)M_{\rm int}$.
In Eq.~(\ref{equa:A}), $\iota$ denotes the angle of inclination of the binary's orbital angular momentum with the line of sight. Since the short gamma ray bursts (SGRBs) are strongly beamed, the binaries should be orientated nearly face on (i.e., $\iota\simeq 0$) as implied by the coincidence observations of SGRBs, and the maximal inclination is about $\iota=20^\circ$.

Once the waveform of the GW is known, the signal-to-noise ratio (SNR) for the network of three independent interferometers can be calculated by
\begin{equation}
\rho=\sqrt{\sum\limits_{i=1}^{3}(\rho^{(i)})^2},
\label{euqa:rho}
\end{equation}
where $\rho^{(i)}=\sqrt{\left\langle \mathcal{H}^{(i)},\mathcal{H}^{(i)}\right\rangle}$, and here the inner product of $a(t)$ and $b(t)$ is defined as \begin{equation}
\left\langle{a,b}\right\rangle=4\int_{f_{\rm lower}}^{f_{\rm upper}}\frac{\tilde a(f)\tilde b^\ast(f)+\tilde a^\ast(f)\tilde b(f)}{2}\frac{df}{S_h(f)},
\label{euqa:product}
\end{equation}
where ``$\sim$" above the function represents the Fourier transform of the each quantity and $S_h(f)$ is the one-side noise power spectral density. Note here that we have taken $S_h(f)$ of the ET to be the same as that in Ref.~\cite{Zhao:2010sz}.

The Fisher information matrix can be used to estimate the instrumental error on the measurement of $d_{L}$,
\begin{align}
\sigma_{d_L}^{\rm inst}\simeq \sqrt{\left\langle\frac{\partial \mathcal H}{\partial d_L},\frac{\partial \mathcal H}{\partial d_L}\right\rangle^{-1}}.
\end{align}
Because $d_{L}$ is independent of other parameters, according to the relation $\mathcal H \propto d_L^{-1}$, we can easily get $\sigma_{d_L}^{\rm inst}\simeq d_L/\rho$. When considering the effect of the inclination angle $\iota$ ($0^{\circ}<\iota<90^{\circ}$), we need to add a factor 2 in front of the error, namely,
\begin{equation}
\sigma_{d_L}^{\rm inst}\simeq \frac{2d_L}{\rho}.
\label{sigmainst}
\end{equation}
In addition, we have to consider the error from weak lensing, which can be expressed as $\sigma_{d_L}^{\rm lens}$ = $0.05z d_L$ \cite{Cai:2016sby}. Therefore, the total error of the luminosity distance is
\begin{align}
\sigma_{d_L}&~~=\sqrt{(\sigma_{d_L}^{\rm inst})^2+(\sigma_{d_L}^{\rm lens})^2} \nonumber\\
            &~~=\sqrt{\left(\frac{2d_L}{\rho}\right)^2+(0.05z d_L)^2}.
\label{sigmadl}
\end{align}

Now, we can generate the catalogue of the GW standard sirens data ($z_{i}$, $d_{L}(z_{i})$, $\sigma_{i}$). In Ref.~\cite{Cai:2016sby}, it pointed out that the constraining capability of about 1000 GW events is similar to that of the Planck mission. Thus, in this paper, we also simulate 1000 GW standard siren data points which are expected to be detected by the ET in its 10-year observation.

In order to constrain the cosmological parameters, we will use the Markov chain Monte Carlo (MCMC) method to infer the posterior probability distributions.
The procedure is as follows. First, we will use the current data combination of CMB+BAO+SN to constrain the I$\Lambda$CDM and I$w$CDM models,
and then we use the obtained best-fit values of the cosmological parameters (except for the coupling constant $\beta$) to simulate the future GW data; due to the central value of the coupling constant $\beta$ being around zero, we take the fiducial value as $\beta=0$ for this parameter.
Next, we will consider the simulated GW standard sirens data in our analysis and combine them with the current cosmological data (i.e., CMB+BAO+SN+GW) to constrain the IDE models again, investigating whether the addition of the GW data can improve the constraints on the parameters of IDE models.

The $\chi^2$ function for the GW observation can be written as
\begin{align}
\chi_{\rm GW}^2=\sum\limits_{i=1}^{N}\left[\frac{\bar{d}_L^i-d_L(\bar{z}_i;\vec{\Omega})}{\bar{\sigma}_{d_L}^i}\right]^2,
\label{equa:chi2}
\end{align}
where $\bar{z}_i$, $\bar{d}_L^i$, and $\bar{\sigma}_{d_L}^i$ are the $i$th redshift, luminosity distance, and error of luminosity distance, respectively, and $\vec{\Omega}$ denotes the set of cosmological parameters.

For the combination of the conventional EM observations and the GW standard siren observation, the total $\chi^{2}_{\rm tot}$ function is
\begin{equation}
\chi^{2}_{\rm tot}=\chi^{2}_{\rm CMB}+\chi^{2}_{\rm BAO}+\chi^{2}_{\rm SN}+\chi^{2}_{\rm GW}.
\end{equation}

\begin{figure*}[!htp]
\includegraphics[scale=0.24]{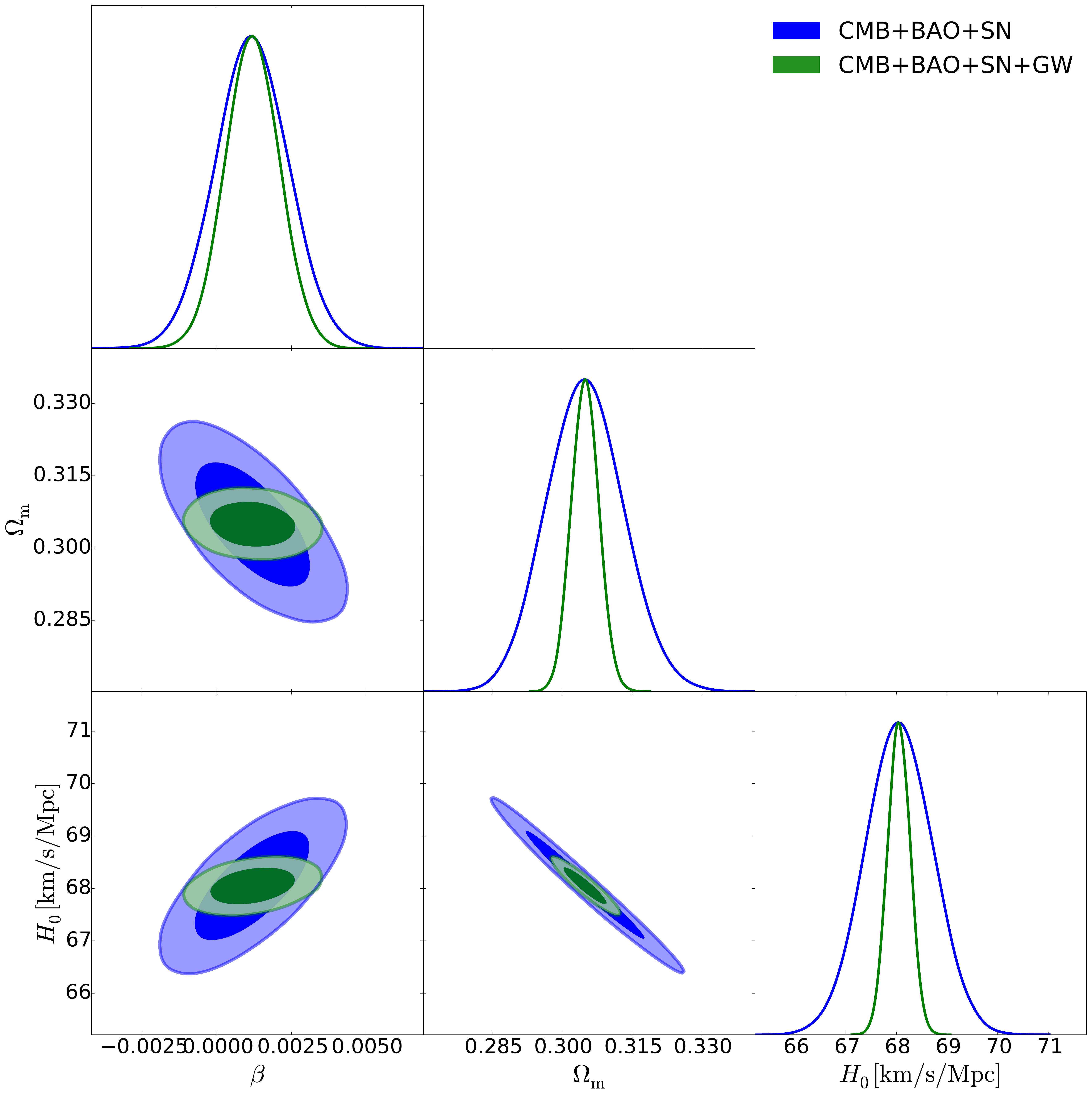}

\centering \caption{\label{fig1} Observational constraints (68.3\% and 95.4\% confidence level) on the I$\Lambda$CDM1 model with $Q=\beta H \rho_{\rm c}$ by using the CMB+BAO+SN and CMB+BAO+SN+GW data.}
\end{figure*}
\begin{figure*}[!htp]
\includegraphics[scale=0.25]{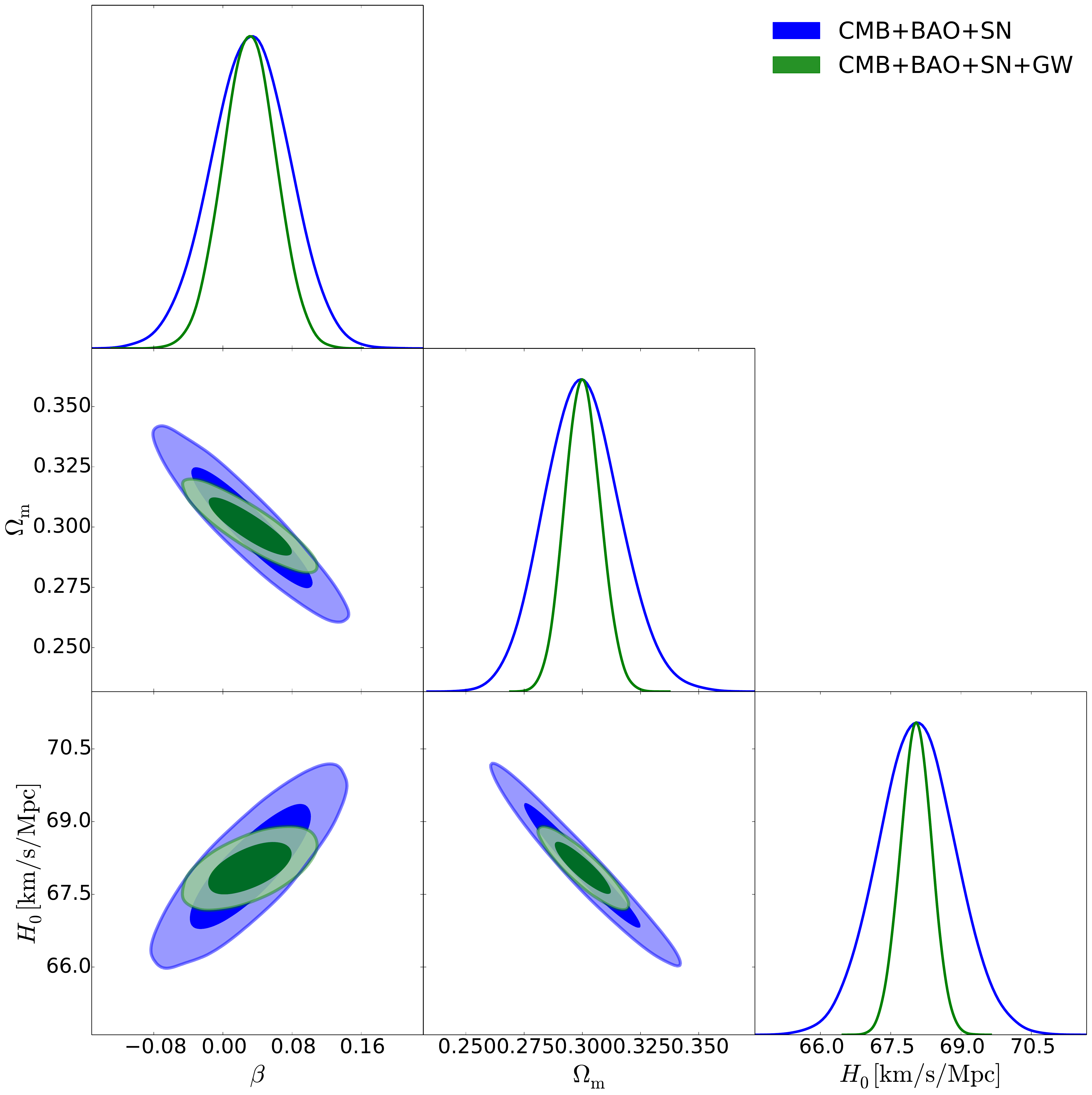}

\centering \caption{\label{fig2} Observational constraints (68.3\% and 95.4\% confidence level) on the I$\Lambda$CDM2 model with $Q=\beta H_{0} \rho_{\rm c}$ by using the CMB+BAO+SN and CMB+BAO+SN+GW data.}
\end{figure*}
\begin{figure*}[!htp]
\includegraphics[scale=0.25]{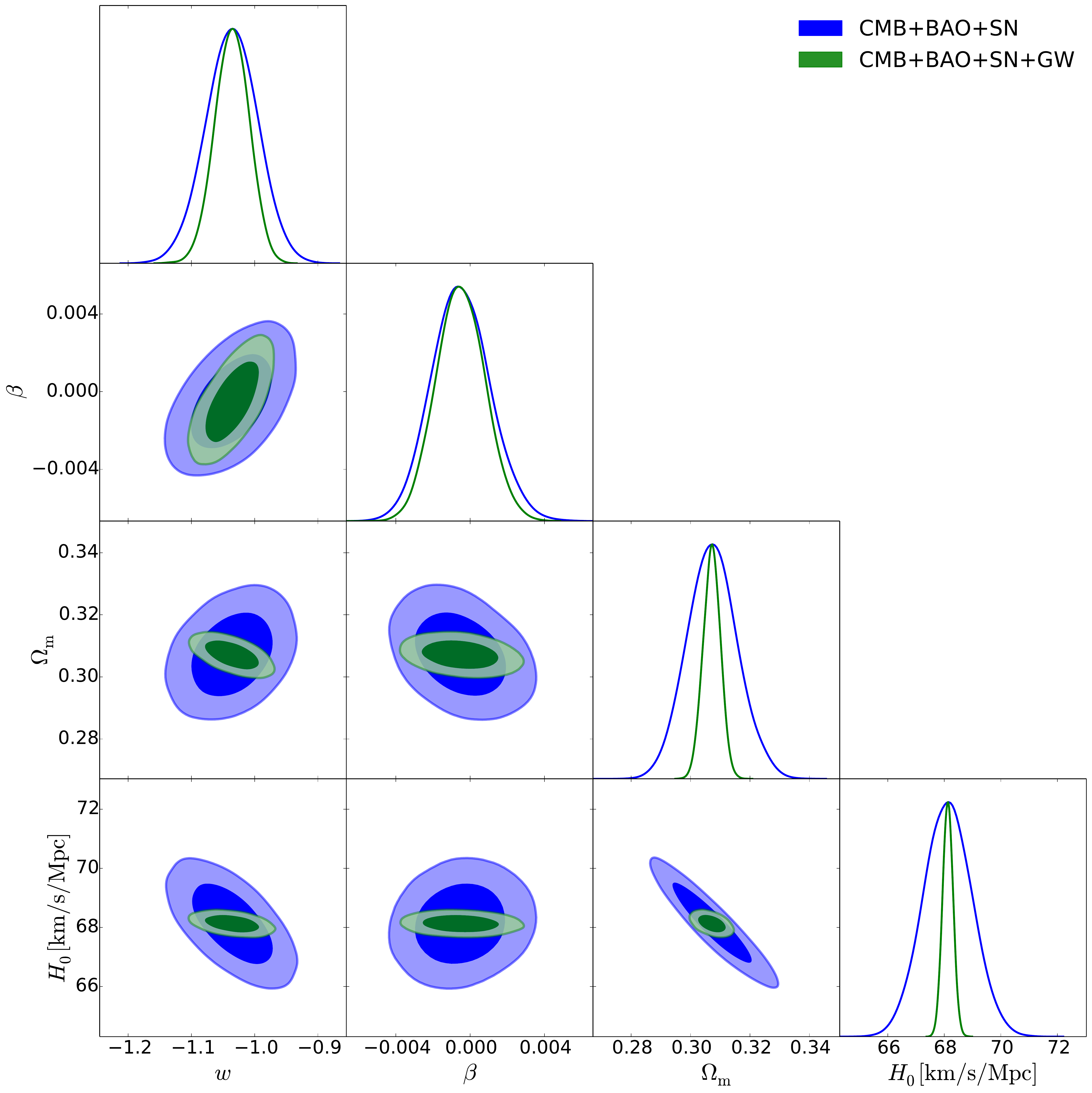}

\centering \caption{\label{fig3} Observational constraints (68.3\% and 95.4\% confidence level) on the I$w$CDM1 model with $Q=\beta H \rho_{\rm c}$ by using the CMB+BAO+SN and CMB+BAO+SN+GW data.}
\end{figure*}

\begin{figure*}[!htp]
\includegraphics[scale=0.25]{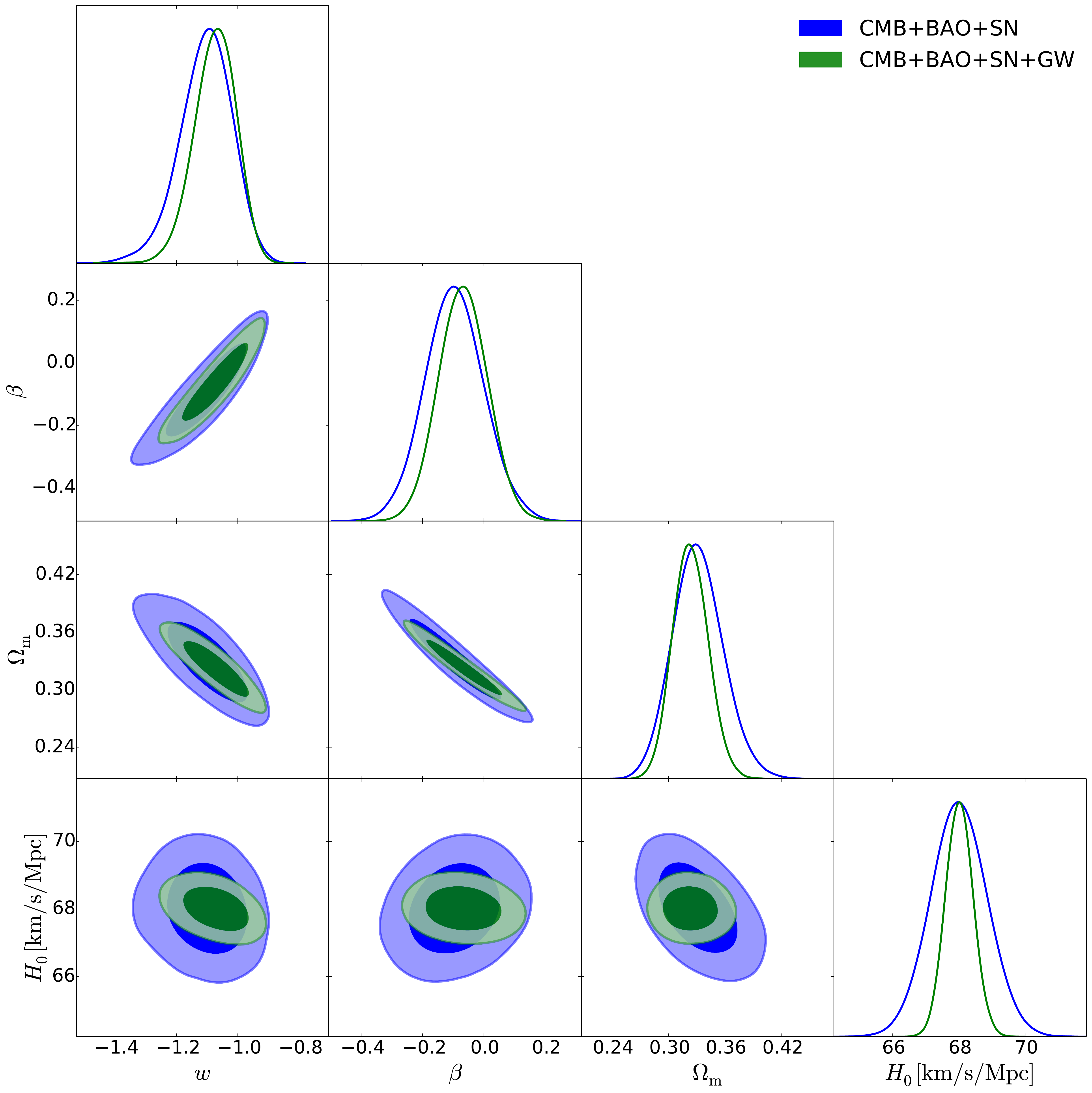}

\centering \caption{\label{fig4} Observational constraints (68.3\% and 95.4\% confidence level) on the I$w$CDM2 model with $Q=\beta H_{0} \rho_{\rm c}$ by using the CMB+BAO+SN and CMB+BAO+SN+GW data.}
\end{figure*}

\begin{figure*}[!htp]
\includegraphics[scale=0.47]{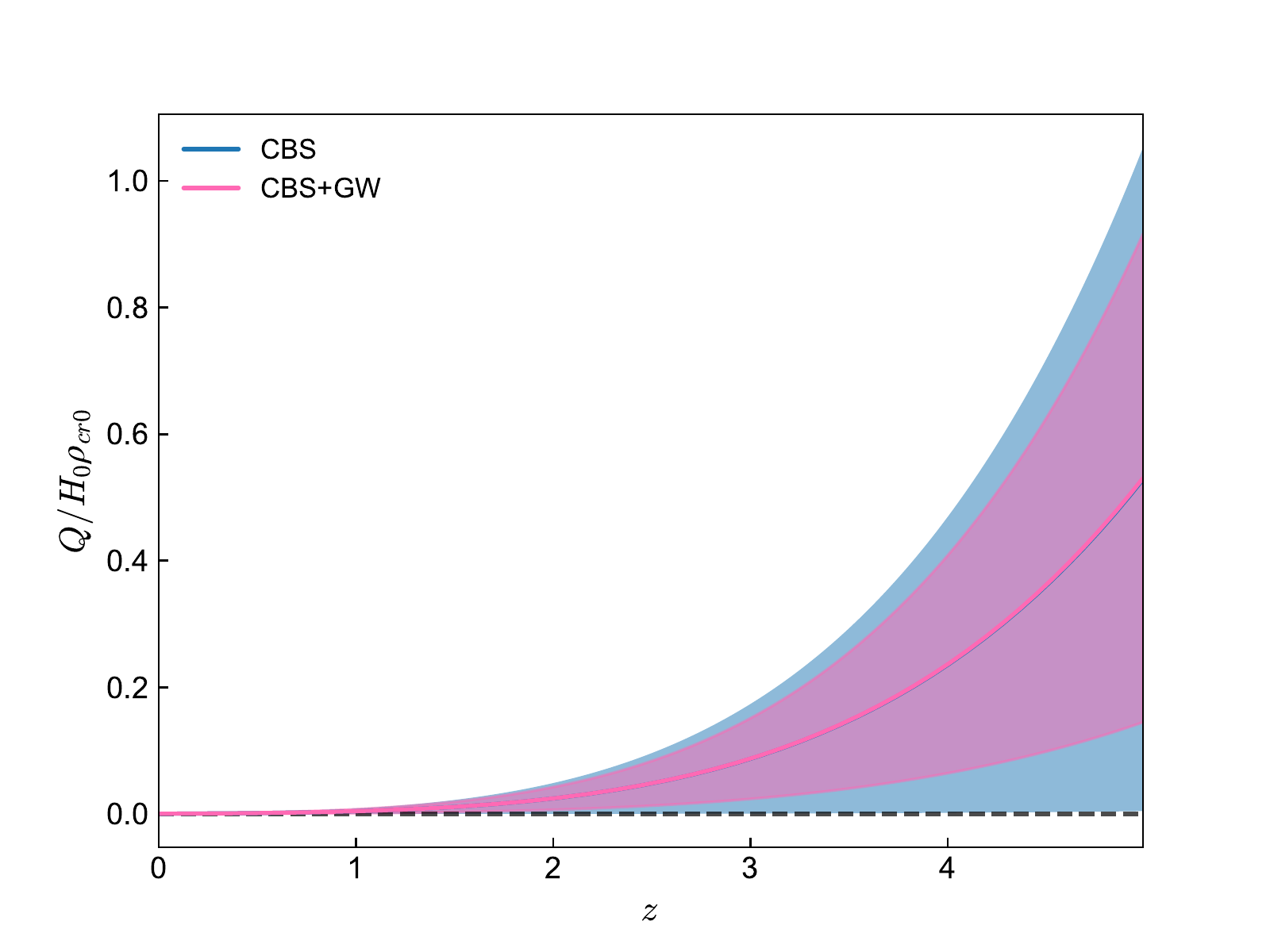}
\includegraphics[scale=0.47]{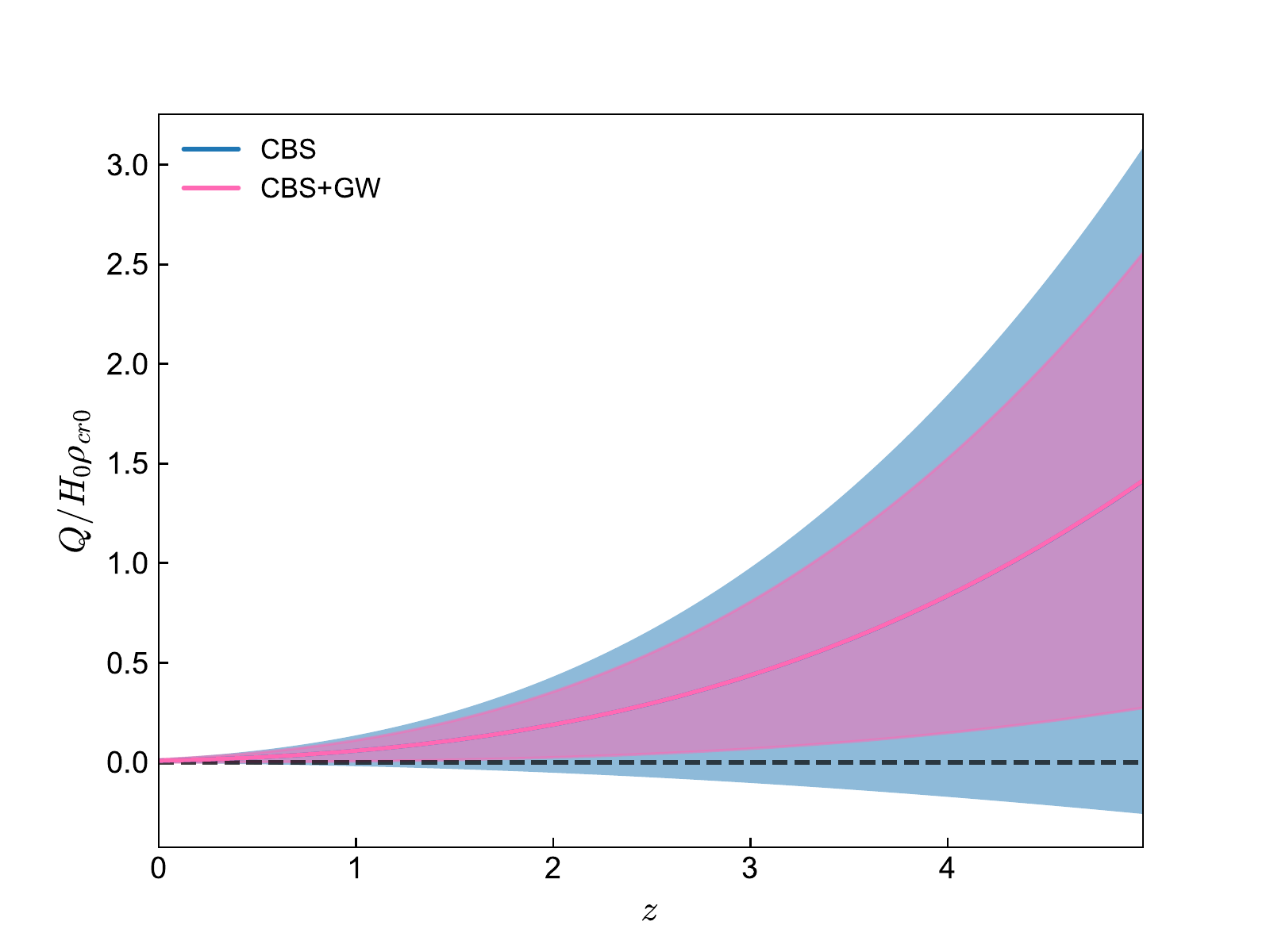}
\centering \caption{\label{fig5} The evolutions of $Q/H_{0}\rho_{\rm cr0}$ (with 1$\sigma$ errors) in the I$\Lambda$CDM1 model with $Q=\beta H \rho_{\rm c}$ (\emph{left}) and in the I$\Lambda$CDM2 model with $Q=\beta H_{0} \rho_{\rm c}$ (\emph{right}), respectively. The black dashed lines denote the noninteracting case ($Q$=0).}
\end{figure*}


\begin{table*}\small
\setlength\tabcolsep{2pt}
\renewcommand{\arraystretch}{1.5}
\centering
\caption{\label{tab1}Fitting results (68.3\% confidence level) for the I$\Lambda$CDM models using CBS and CBS+GW. Here, CBS stands for CMB+BAO+SN.}
\begin{tabular}{ccccccccc}

\hline Model &\multicolumn{2}{c}{I$\Lambda$CDM1 ($Q=\beta H\rho_{\rm c}$)}&&\multicolumn{2}{c}{I$\Lambda$CDM2 ($Q=\beta H_{0}\rho_{\rm c}$)}\\
           \cline{2-3}\cline{5-6}
       Data  & CBS&CBS+GW &&CBS&CBS+GW \\

\hline
$\Omega_{\rm m}$                         &$0.3050^{+0.0080}_{-0.0081}$
                                         &$0.3050\pm0.0029$
                                         &&$0.3000^{+0.0150}_{-0.0170}$
                                         &$0.3002^{+0.0075}_{-0.0074}$
                                         \\

$H_0\,[{\rm km}/{\rm s}/{\rm Mpc}]$      &$68.05^{+0.65}_{-0.64}$
                                         &$68.05\pm0.22$
                                         &&$68.06\pm0.80$
                                         &$68.04\pm0.33$
                                         \\

$\beta$                                  &$0.00120\pm0.00120$
                                         &$0.00120\pm0.00088$
                                         &&$0.03100\pm0.04400$
                                         &$0.03100\pm0.03000$
                                         \\

\hline
\end{tabular}

\end{table*}
\begin{table*}\small
\setlength\tabcolsep{2pt}
\renewcommand{\arraystretch}{1.5}
\centering
\caption{\label{tab2}Fitting results (68.3\% confidence level) for the I$w$CDM models using CBS and CBS+GW. Here, CBS stands for CMB+BAO+SN.}
\begin{tabular}{ccccccccc}

\hline Model &\multicolumn{2}{c}{I$w$CDM1 ($Q=\beta H\rho_{\rm c}$)}&&\multicolumn{2}{c}{I$w$CDM2 ($Q=\beta H_{0}\rho_{\rm c}$)}\\
           \cline{2-3}\cline{5-6}
       Data  & CBS&CBS+GW &&CBS&CBS+GW\\

\hline
$\Omega_{\rm m}$                         &$0.3073^{+0.0081}_{-0.0082}$
                                         &$0.3071^{+0.0028}_{-0.0029}$
                                         &&$0.3320^{+0.0250}_{-0.0280}$
                                         &$0.3240^{+0.0180}_{-0.0200}$\\

$H_0\,[{\rm km}/{\rm s}/{\rm Mpc}]$      &$68.13^{+0.84}_{-0.83}$
                                         &$68.13\pm0.18$
                                         &&$68.00\pm0.82$
                                         &$68.02\pm0.40$\\

$\beta$                                  &$-0.0005\pm0.0015$
                                         &$-0.0005\pm0.0013$
                                         &&$-0.0950\pm0.0930$
                                         &$-0.0670\pm0.0780$\\

$w$                                      &$-1.036\pm0.040$
                                         &$-1.036\pm0.026$
                                         &&$-1.105^{+0.093}_{-0.075}$
                                         &$-1.075^{+0.074}_{-0.062}$\\

\hline
\end{tabular}

\end{table*}
\begin{table*}\small
\setlength\tabcolsep{2pt}
\renewcommand{\arraystretch}{1.5}
\centering
\caption{\label{tab3}Constraint errors for cosmological parameters of the I$\Lambda$CDM models and the I$w$CDM models using CBS and CBS+GW. Here, CBS stands for CMB+BAO+SN.}
\begin{tabular}{ccccccccccc}

\hline Model &\multicolumn{2}{c}{I$\Lambda$CDM1 ($Q=\beta H\rho_{\rm c}$)}&\multicolumn{2}{c}{I$\Lambda$CDM2 ($Q=\beta H_{0}\rho_{\rm c}$)}&&\multicolumn{2}{c}{I$w$CDM1 ($Q=\beta H\rho_{\rm c}$)}&\multicolumn{2}{c}{I$w$CDM2 ($Q=\beta H_{0}\rho_{\rm c}$)}\\
           \cline{2-5}\cline{7-10}
       Data  & CBS&CBS+GW &CBS&CBS+GW &&CBS&CBS+GW&CBS&CBS+GW\\

\hline
$\sigma(\Omega_{\rm m})$                         &$0.0081$
                                         &$0.0029$
                                         &$0.0160$
                                         &$0.0075$
                                         &&$0.0082$
                                         &$0.0029$
                                         &$0.0265$
                                         &$0.0190$\\

$\sigma(H_0)$      &$0.645$
                                         &$0.220$
                                         &$0.800$
                                         &$0.330$
                                         &&$0.835$
                                         &$0.180$
                                         &$0.820$
                                         &$0.400$\\

$\sigma(\beta)$                                  &$0.00120$
                                         &$0.00088$
                                         &$0.04400$
                                         &$0.03000$
                                         &&$0.00150$
                                         &$0.00130$
                                         &$0.09300$
                                         &$0.07800$\\

$\sigma(w)$                                      &$-$
                                         &$-$
                                         &$-$
                                         &$-$
                                         &&$0.0400$
                                         &$0.0260$
                                         &$0.0845$
                                         &$0.0683$\\

\hline
\end{tabular}

\end{table*}
\begin{table*}\small
\setlength\tabcolsep{2pt}
\renewcommand{\arraystretch}{1.5}
\centering
\caption{\label{tab4}Constraint accuracies for cosmological parameters of the I$\Lambda$CDM models and the I$w$CDM models using CBS, and CBS+GW. Here, CBS stands for CMB+BAO+SN.}
\begin{tabular}{ccccccccccc}

\hline Model &\multicolumn{2}{c}{I$\Lambda$CDM1 ($Q=\beta H\rho_{\rm c}$)}&\multicolumn{2}{c}{I$\Lambda$CDM2 ($Q=\beta H_{0}\rho_{\rm c}$)}&&\multicolumn{2}{c}{I$w$CDM1 ($Q=\beta H\rho_{\rm c}$)}&\multicolumn{2}{c}{I$w$CDM2 ($Q=\beta H_{0}\rho_{\rm c}$)}\\
           \cline{2-5}\cline{7-10}
       Data  & CBS&CBS+GW &CBS&CBS+GW &&CBS&CBS+GW&CBS&CBS+GW\\

\hline
$\varepsilon(\Omega_{\rm m})$                         &$0.0266$
                                         &$0.0095$
                                         &$0.0533$
                                         &$0.0250$
                                         &&$0.0267$
                                         &$0.0094$
                                         &$0.0798$
                                         &$0.0586$\\

$\varepsilon(H_0)$      &$0.0095$
                                         &$0.0032$
                                         &$0.0118$
                                         &$0.0049$
                                         &&$0.0123$
                                         &$0.0026$
                                         &$0.0121$
                                         &$0.0059$\\


$\varepsilon(w)$                                      &$-$
                                         &$-$
                                         &$-$
                                         &$-$
                                         &&$0.0386$
                                         &$0.0251$
                                         &$0.0765$
                                         &$0.0635$\\

\hline
\end{tabular}

\end{table*}

\section{Results and discussion}\label{sec4}

The main constraint results ate summarized in Figs.~\ref{fig1}--\ref{fig4} and Tables~\ref{tab1}--\ref{tab4}. In Figs.~\ref{fig1}--\ref{fig4}, the constraint results for the I$\Lambda$CDM1 model with $Q=\beta H\rho_{\rm c}$, the I$\Lambda$CDM2 model with $Q=\beta H_{0}\rho_{\rm c}$, the I$w$CDM1 model with $Q=\beta H\rho_{\rm c}$, and the I$w$CDM2 model with $Q=\beta H_{0}\rho_{\rm c}$ are shown, respectively. In these figures, one-dimensional marginalized posterior distributions and the two-dimensional contours (68.3\% and 95.4\% confidence level) from CMB+BAO+SN and CMB+BAO+SN+GW are colored by blue and green, respectively. The fit values of the cosmological parameters for the IDE models are given in Tables \ref{tab1} and \ref{tab2}. The constraint errors of the cosmological parameters are given in Table \ref{tab3}, and the constraint accuracies are given in Table \ref{tab4}. Here, for a parameter $\xi$, the accuracy $\varepsilon(\xi)$ is defined as $\varepsilon(\xi)=\sigma(\xi)/\xi$, which is the relative error. For convenience, the data combination ``CMB+BAO+SN" is also abbreviated as ``CBS" in the following.

At first glance, we can easily find that the addition of the GW standard siren data can tighten the constraint on $H_0$ and $\Omega_{\rm m}$ significantly (except the case in the I$w$CDM2 model with $Q=\beta H_{0}\rho_{\rm c}$, which will be discussed in the following), and for the other parameters the constraints are slightly weaker.

The constraint results for the I$\Lambda$CDM1 model with $Q=\beta H\rho_{\rm c}$ are shown in Fig.~\ref{fig1}. We find that the CBS data provide a 0.95\% measurement for $H_0$, whereas the combined CBS+GW data provide a 0.32\% measurement. For the parameter $\Omega_{\rm m}$, the CBS data can give a constraint accuracy of 2.66\%. When adding the GW data, the constraint accuracy for $\Omega_{\rm m}$ is improved to 0.95\% (using the CBS+GW data). Obviously, both parameters can be constrained more stringent with the help of the GW data. Note here that since the central value of the coupling constant $\beta$ in IDE models is around zero, the relative error for this parameter will be immensely influenced by the statistic fluctuations. Therefore, the absolute error is more reliable for quantifying the improvement for constraining this parameter. The addition of the GW data will tighten the constraint on the coupling constant $\beta$, with the absolute error improved from $\sigma(\beta)=1.2 \times 10^{-3}$ to $\sigma(\beta)=8.8 \times 10^{-4}$.

The constraint results for the I$\Lambda$CDM2 model with $Q=\beta H_{0}\rho_{\rm c}$ are shown in Fig.~\ref{fig2}. For this model, the CBS data can provide a 1.18\% measurement for $H_0$, while the CBS+GW data can measure $H_0$ at the 0.49\% level. With respect to the parameter $\Omega_{\rm m}$, the CBS+GW data can give a 2.50\% constraint accuracy on $\Omega_{\rm m}$, better than the constraint using the CBS data of a 5.33\% accuracy. For the coupling constant $\beta$, the result is similar to the case of the I$\Lambda$CDM1 model, and the constraint error could be improved  from $\sigma(\beta)=4.4 \times 10^{-2}$ to $\sigma(\beta)=3 \times 10^{-2}$ with the addition of the GW data.

The results for the I$w$CDM1 model with $Q=\beta H\rho_{\rm c}$ are shown in Fig.~\ref{fig3}. From this figure, we find that the situation is similar to that of the I$\Lambda$CDM models. We can see that the CBS data can only provide a 1.23\% measurement for $H_0$, while the combined CBS+GW data constrain $H_0$ to a 0.26\% accuracy. As for the measurement of $\Omega_{\rm m}$, we find that the constraint result using the CBS+GW data is also better than that using the CBS data. When adding the GW data, the constraint accuracy of $\Omega_{\rm m}$ will be improved from 2.67\% to 0.94\%. For the parameter $w$, there is a slight improvement when adding the GW data, with the accuracy enhanced from 3.86\% to 2.51\%. For the coupling parameter $\beta$, the constraint errors are $\sigma(\beta)=1.5 \times 10^{-3}$ and $\sigma(\beta)=1.3 \times 10^{-3}$ using CBS and CBS+GW, respectively. The improvement in this case is not evident.

Finally, we investigate the I$w$CDM2 model with $Q=\beta H_{0}\rho_{\rm c}$, of which the constraint results are shown in Fig.~\ref{fig4}. We find that the constraint accuracy on $\Omega_{\rm m}$ is worse compared with the cases in the above three models. When adding the GW data, the constraint on $\Omega_{\rm m}$ is at a 5.86\% accuracy (using CBS+GW data), slightly better than that using the CBS data at a 7.98\% accuracy. In addition, we also find that the CBS data provide a 1.21\% measurement for $H_0$, and the combined CBS+GW data provides a 0.59\% measurement for $H_0$. Similar to the case of I$w$CDM1 model with $Q=\beta H\rho_{\rm c}$, the accuracy of $w$ is only slightly improved, from 7.65\% to 6.35\%. For the coupling parameter $\beta$, when we add the GW data, the constraint error is slightly decreased, from $\sigma(\beta)=9.3 \times 10^{-2}$ to $\sigma(\beta)=7.8 \times 10^{-2}$.

In addition, we also plot the reconstructed evolutions of the interaction term $Q(z)$ in the I$\Lambda$CDM1 and I$\Lambda$CDM2 models in Fig.~\ref{fig5}. Here we show $Q/H_{0}\rho_{\rm cr0}$ versus $z$, where $\rho_{\rm cr0}=3M_{\rm pl}^{2} H_{0}^{2}$ is the present-day critical density of the universe. The best-fit line and the 1$\sigma$ region are shown in the figure, and the two panels are for the I$\Lambda$CDM1 and I$\Lambda$CDM2 models, respectively. It is clear to see that the addition of the GW data could significantly shrink the uncertainty in the reconstruction of the interaction term.

Note here that, as an another third-generation ground-based GW observatory in addition to the ET (in the Europe), the Cosmic Explorer (CE) has also been proposed to be built in the United States. This project is scheduled to start its observation in the mid-2030s. Some forecast studies on constraining cosmological parameters using the GW standard sirens observed by the CE have been made in the literature. The cosmological parameter constraining capability of the CE is slightly better than that of the CE, as shown in Refs.~\cite{Jin:2020hmc,Zhao:2017cbb}.


In summary, for all the IDE models considered in this paper, the future GW standard siren data observed by the ET can indeed improve the constraint accuracies of cosmological parameters (e.g., $\Omega_{\rm m}$, $H_0$, and $w$). Specifically, for the coupling parameter $\beta$, when adding the GW data, the constraint error is also evidently decreased.

\section{Conclusion}\label{sec5}

In this paper, we have investigated                                                                                                                                                                                                                                                                                                                                                                                                                                                                                                                                                                                                                                                                                                                                                                                                                                                                                                                                                                                                                                                                                                                                                                                                                                                                                                                                                                                                                                                                                                                                                                                                                                                                                                                                                                                                                                                                                                                                                                                                                                                                                                                                                                                                                                                        how the future GW standard sirens observed by the next-generation ground-based GW detectors would impact on the cosmological parameter estimation for the IDE models. We consider four typical IDE models, i.e., the I$\Lambda$CDM1 ($Q=\beta H\rho_{\rm c}$) model, the I$\Lambda$CDM2 ($Q=\beta H_{0}\rho_{\rm c}$) model, the I$w$CDM1 ($Q=\beta H\rho_{\rm c}$) model, and the I$w$CDM2 ($Q=\beta H_{0}\rho_{\rm c}$) model. To study the impacts of the GW data, we also consider the conventional EM observations in this paper including the Planck 2015 CMB data, the BAO measurements, and the SN data of Pantheon compilation. 
For the GW data, we simulate 1000 GW multi-messenger events based on the ET's ten-year observation. In order to quantify the constraint capability of the GW data, we consider two data combinations, i.e., CBS and CBS+GW, to constrain the cosmological models.

We find that the future GW standard sirens can significantly improve the constraints on most of the cosmological parameters for all the IDE models. When adding the GW standard siren data, the constraint accuracy of $H_0$ can be remarkably improved, from 0.95\%, 1.18\%, 1.23\%, and 1.21\% to 0.32\%, 0.49\%, 0.26\%, and 0.59\% for the I$\Lambda$CDM1, I$\Lambda$CDM2, I$w$CDM1, and I$w$CDM2 models, respectively. Moreover, as for the parameter $\Omega_{\rm m}$, the constraint accuracy is improved from 2.66\%, 5.33\%, 2.67\%, and 7.98\% to 0.95\%, 2.50\%, 0.94\%, and 5.86\%, for the four considered models, respectively. For the coupling constant $\beta$, when adding the GW data, the constraint error $\sigma(\beta)$ can also be decreased, from $1.2 \times 10^{-3}$, $4.4 \times 10^{-2}$, and $9.3 \times 10^{-2}$ to $8.8 \times 10^{-4}$, $3.0 \times 10^{-2}$, and $7.8 \times 10^{-2}$ for the I$\Lambda$CDM1, I$\Lambda$CDM2, and I$w$CDM2 models, respectively. While for the I$w$CDM1 model, the improvement is not so evident for this parameter, from $\sigma(\beta)=1.5 \times 10^{-3}$ to $\sigma(\beta)=1.3 \times 10^{-3}$. For the parameter $w$ in the I$w$CDM models, the constraint accuracy could also be improved when adding the GW data, from 3.86\% to 2.51\% for the I$w$CDM1 model and from 7.65\% to 6.35\% for the I$w$CDM2 model.

It is shown in this paper that for the IDE models the constraint precisions for cosmological parameters can be promoted effectively with the consideration of future GW observations. The results presented here are consistent with the previous studies on other dark energy models. We conclude that the GW standard sirens would be developed into a powerful cosmological probe in the future. Due to the fact that the GW observations can measure the absolute distance scale in cosmology, the standard sirens can be used to break the cosmological parameter degeneracies inherent in the conventional EM observations. The next-generation ground-based GW detectors and the future space-based GW detectors would observe a large number of GW events in multi-frequency bands, providing a large sample of standard sirens that will greatly promote the development of cosmology. We need more detailed studies on the standard siren cosmology.


\begin{acknowledgments}

We thank Jing-Zhao Qi, Ze-Wei Zhao, and Ling-Feng Wang for helpful discussions.
This work was supported by the National Natural Science Foundation of China (Grants Nos.~11975072, 11875102, 11835009, and 11690021), the Liaoning Revitalization Talents Program (Grant No. XLYC1905011), and the Fundamental Research Funds for the Central Universities (Grant No. N2005030).

\end{acknowledgments}

\end{document}